\documentclass[3p,times]{elsarticle}

\usepackage{ecrc}
\usepackage{amssymb}   
\usepackage{amsmath}   
\usepackage{tensor}
\usepackage{multicol}

\volume{00}

\firstpage{1}

\journalname{Nuclear Physics A}

\runauth{A. Rothkopf}

\jid{nupha}

\usepackage{amssymb}

\usepackage[figuresright]{rotating}

\usepackage{slashed}

\begin{document}

\begin{frontmatter}



\dochead{}

\title{Heavy Quarkonium in the Quark Gluon Plasma from Effective Field Theories and Potentials}


\author{A. Rothkopf}

\address{Albert Einstein Center for Fundamental Physics, Institute for Theoretical Physics, University of Bern, Sidlerstrasse 5, CH-3012 Bern, Switzerland}

\begin{abstract}
The measurements of heavy quarkonium suppression at RHIC and LHC urge theory to develop intuitive as well as quantitative methods for the description of $Q\bar{Q}$ melting in the quark-gluon plasma. Here I will present a brief sketch on the effective field theory strategies underlying the definition of the heavy quark static potential and report on two recent advances in the extraction and interpretation of such a potential.
On the one side, progress has been made in obtaining its values from lattice QCD, which promises to make possible investigating its real and imaginary part non-perturbatively. On the other side, the existence of an imaginary part emphasizes the dynamical nature of the melting process and invites us to make a direct connection to the framework of open quantum systems.
\end{abstract}

\begin{keyword}
heavy quarkonium \sep effective field theory \sep in-medium static potential \sep open quantum system

\end{keyword}

\end{frontmatter}


\section{Introduction}
The past decade has seen a golden age in the study of relativistic heavy-ion collision. After the inauguration of the RHIC facility at the Brookhaven National Laboratory in 2000, experimental collaborations such as PHENIX and STAR were able to conclusively deduce the existence of a deconfined phase of strongly interacting matter. In collision of gold nuclei at a center of mass energy per nucleon of $\sqrt{s_{NN}}=200GeV$, the occurrence of e.g. collective (elliptic) flow \cite{Afanasiev:2009wq,Adamczyk:2011aa} and quenching of jet structures\cite{Adler:2002tq,Adler:2005ee} provided unambiguous signals for the presence of a quark-gluon plasma (QGP).

Another candidate for a clean signal of deconfinement, introduced in the seminal work by Matsui and Satz \cite{Matsui:1986dk}, is the suppression of heavy quark bound states. Initially interest focussed on the melting of charmonium, the ground state of the $c\bar{c}$ vector channel. Even though the process of melting is amenable to analytic and numerical treatment as we will discuss later on and suppression of heavy quarkonium has been observed\cite{Adare:2006ns,Tang:2011kr,Chatrchyan:2012np,Abelev:2012rv}, it has become clear that the physics of $c\bar{c}$ in relativistic heavy ion collisions is quite complex. Cold nuclear matter effects (observed already in d+Au collisions) as well as final state effects (e.g. recombination and feed-down) lead to a competition between suppression and replenishment in the occupation of the ground state that is not necessarily related to the intrinsic stability, i.e. melting of the two-body system itself.

The response of bound states to the presence of a thermalized hot medium can be investigated more cleanly once the next heavier quark flavor $b$ and $\bar{b}$ is produced in adequate numbers in collisions. The necessary energies to do so are available at the Large Hadron Collider (LHC) at CERN, which recently commenced its first heavy ion runs, operating at a $\sqrt{s_{NN}}=2.76TeV$. A di-muon spectrum showing clean signals for ground and two excited states in $p+p$ and melting of the higher lying states in $Pb+Pb$ collisions was published last year by the CMS collaboration \cite{Chatrchyan:2011pe}. Since $b\bar{b}$ is produced only in limited numbers in the hard processes in the initial phase of the collision, recombination is much less pronounced than for charmonium. Most importantly however, since the top quark is comparatively heavy, there will not be any appreciable feed down from $t\bar{t}$ bound states. This in turn should allow to establish a closer connection between bottomonium spectra and the 
idealized theoretical descriptions of heavy quarkonium melting.

Several strategies, varying in their degree of sophistication, have been developed to model the suppression of heavy quarkonia in relativistic heavy ion collisions (see e.g. \cite{Strickland:2011mw,Uphoff:2012gb,Zhao:2010nk,Brezinski:2011ju,Song:2011nu}). In the following I will solely focus on the subset of the problem that is concerned with the melting of the bound state in an otherwise fully thermalized heat bath. As we will see, a dynamical description is possible within a non-relativistic framework based on effective field theories, which ultimately leads to a potential for a two-body wave-function in a Schroedinger-type equation.

Past attempts to describe the stability of heavy quarkonium often relied on a static notion of a model potential and the bound states it can support. Since the early works of \cite{Nadkarni:1986cz,Nadkarni:1986as} the color singlet free energies $F^1(r)$ and subsequently the color singlet internal energies $U^1(r)$ or even linear combination of both quantities were identified with the potential acting between a static $Q$ and $\bar{Q}$ (see e.g. \cite{Satz:2005hx}). All of these quantities can be obtained from a suitable correlation function in lattice QCD, which is evaluated at a single imaginary time step $\tau=\beta$. Be aware that these purely real potentials are model choices, since there is no derivation available that connects quantities such as $F^1(r)$ to the real-time evolution of heavy quarkonium in a Schroedinger equation. 

Even more interestingly, it has been shown in resummed perturbation theory that the potential in the QGP in general does not only contain a real part but in addition features an imaginary part \cite{Laine:2006ns,Beraudo:2007ky}. Hence the notion of a well defined bound state above the deconfinement temperature appears inapplicable and the question of stability and melting need to be formulated in a truly dynamical fashion.

We wish to show in the following how the Schroedinger equation for $Q\bar{Q}$ arises from a systematic coarse graining procedure, starting from QCD and how the values of the potential can be extracted non-perturbatively from lattice QCD. At last we will briefly touch upon our recent proposal on how to interpret such a, in general complex valued, potential as a manifestation of the open quantum systems nature of the heavy quark bound state.\vspace{-0.3cm}

\section{The Effective Field Theory Strategy}

A characteristic feature of heavy quarkonium is the presence of a wide separation of intrinsic scales. In the case of bottomonium e.g., the rest mass of the constituent heavy quarks $m_b\simeq4.6GeV$ is significantly larger than both the dynamical scale of QCD $\Lambda_{QCD}\sim200MeV$ and the temperatures around the deconfinement phase transition $T_C\sim200MeV$. This tells us that neither quantum fluctuations, nor thermal fluctuations are likely to excite a $Q\bar{Q}$ pair from the medium. If the velocity of the constituent quarks are also small, a hierarchy emerges from the hard via a soft to the ultra-soft scale\vspace{-0.3cm}
\begin{align}
 m_Q\gg m_Qv \gg m_Qv^2
\end{align}
along which appropriate effective field theory descriptions can be constructed \cite{Brambilla:1999xf,Brambilla:2004jw,Brambilla:2008cx}. The term appropriate means that we have to choose a set of relevant degrees of freedom at each energy scale and determine the most general action for these fields, which is compatible with the symmetries of the underlying field theory. 

\subsection{NRQCD: Heavy Quarks as Pauli Spinor Fields}

The first reduction in the complexity of the problem is achieved by treating explicitly only the physics of heavy quarks and the medium below the hard scale.  The process of integrating out the higher energies from  the underlying theory of QCD with its matrix valued gauge fields $A^\mu=A^\mu_aT^a$, light medium quarks $q$ and the heavy quark field $Q$\vspace{-0.2cm}
\begin{align}
 {\cal L}_{\rm QCD} = -\frac{1}{4} F^{\mu\nu a}F_{\mu\nu a} + \sum_{l=1}^{N_f}\bar{q}^l\Big(i\gamma^\mu (\partial_\mu + igA_\mu) -m^l_q\Big) q^l + \bar{Q} (i\gamma^\mu (\partial_\mu + igA_\mu) -m_Q\Big) Q,\label{QCDLag}
\end{align}
leads us to the theory of NRQCD. The first step to take lies in identifying the degrees of freedom to use. 

In case of the heavy quarks, pair production of $Q$ and $\bar{Q}$ cannot persist below the threshold $2m_Q$. Hence it seems advantageous to use the upper $\xi$ and lower components $\chi$ of the original Dirac four spinor $Q=(\xi,\chi)$. Fortunately the problem of separating upper and lower components can be treated systematically by use of the Foldy-Tani-Wouthuysen transformation, which amounts to a systematic expansion of the heavy quark part of the QCD action in a series involving terms of increasing powers of $m_Q^{-1}$. The first few terms in the effective Lagrangian read \cite{Brambilla:2004jw}:\vspace{-0.2cm}
\begin{align}
 {\cal L}_{\rm NRQCD} = -\frac{1}{4} F^{\mu\nu}_aF_{\mu\nu}^a + \sum_{l=1}^{N_f}\bar{q}^l (\slashed{D}-m_q) q^l+\xi^\dagger(iD_0-m_Q+\frac{c_1}{2m_Q}{\bf D}^2+\ldots)\xi + \chi^\dagger(iD_0-m_Q+\frac{c_1}{2m_Q}{\bf D}^2+\ldots)^\dagger\chi\label{NRQCDLag}
\end{align}
Note that each term featuring an explicit factor of $m_Q^{-1}$ is multiplied by a so called Wilson coefficient $c_i$. This in general complex number encodes the remnant effects of the physics at the higher energy scales, which are not treated explicitly anymore. E.g. the annihilation of $Q$ and $\bar{Q}$ into hard gluons will lead to an imaginary part in these coefficients, since the hard gluons themselves are not part of NRQCD\footnote{Since in NRQCD we integrate out all medium degrees of freedom $A^\mu$, $q$, $\bar{q}$ with four momentum $\sim m_Q$ we have to expect that correction terms also for the gluonic and light quark part of the QCD action will arise.}. For details on the determination of the $c_i$'s by a comparison of correlation functions in QCD and NRQCD see \cite{Brambilla:2004jw}. Since in the following we are interested only in deriving the static and spin-independent potential we treat them as set to unity in the remainder of this text.

The expansion in $m_Q^{-1}$ is systematic in the sense that after an inclusion of an infinite number of terms, the original QCD Lagrangian is recovered. It is hence unavoidable that at some order couplings between $\xi$ and $\chi$ fields resurface, which in turn void the assumption of an absence of pair production. Already at order $m_Q^{-2}$ one finds terms such as $\xi^\dagger\chi\chi^\dagger\xi$ or $\xi^\dagger T^a\chi\chi^\dagger T^a\xi$. Thus any non-relativistic potential picture obtained through NRQCD is limited to orders up to $m_Q^{-2}$ in this expansion.

Even though our aim is to construct a potential picture for heavy quarkonium, there is no mention of such concept at this point. In particular the quark and anti-quark are still treated as fields with the interaction being mediated by the soft gluon field. In order to go over to a truly non-relativistic language, i.e. particles propagating according to a Schroedinger equation with a potential $V(r)$, we can continue and integrate out the soft scale for the medium degrees of freedom.  

At this point we encounter the possibility to choose two differently nuanced points of view. Either we can focus on the bound state nature of the two-body system or instead emphasize the individual nature of the two separate quarks. The field theory of pNRQCD \cite{Brambilla:1999xf} follows the former, the quantum mechanical path integral approach \cite{Barchielli:1986zs,Barchielli:1988zp} follows the latter route.

\subsection{pNRQCD: Heavy Quark Pairs as Singlet and Octet fields}

Let us start with pNRQCD \cite{Brambilla:1999xf}, which is constructed again in a threefold way: determine the relevant degrees of freedom, construct the most general Lagrangian and obtain the new set of Wilson coefficients by matching, i.e. calculating the same correlation functions in (NR)QCD and pNRQCD and setting them equal at a chosen scale.

The quark and anti-quark color quantum numbers allow $Q\bar{Q}$ to form either a singlet or octet. Thus one chooses corresponding matrix valued fields $S(R,r,t)=\mathbf{1}_c S(R,r,t)$ and $O(R,r,t)=T^aO^a(R,r,t)$ that characterize the two-body configuration ($R=\frac{1}{2}|{\mathbf x}+{\mathbf y}|$,$r=|{\mathbf x}-{\mathbf y}|$) of the heavy $Q({\mathbf x},t)\bar{Q}({\mathbf y},t)$. Since soft gluons are integrated out in pNRQCD, their contribution to the interaction between the heavy quarks appears in the Lagrangian as an instantaneous potential. Ultrasoft gluons at the scale $E_{us}\sim mv^2$ on the other hand are still explicit degrees of freedom and their fields will couple to the singlet and octet states. To make explicit the low-momentum nature of the remaining electric and magnetic color fields, they are usually multipole expanded in the relative coordinate $r$. The pNRQCD Lagrangian thus reads at first order in $r$ \cite{Brambilla:1999xf}
\begin{align}
&{\cal L}_{\rm pNRQCD}= Tr\Bigg\{ S^\dagger \Big(i\partial_0-2m_Q-V_s(r)-\frac{1}{m_Q}V^{(1)}_s(r)-\ldots\Big)S + O^\dagger \Big(i\partial_0-2m_Q-V_o(r)-\frac{1}{m_Q}V^{(1)}_o(r)-\ldots\Big)O\Bigg\}\label{pNRQCDLag}\\
\nonumber+&gV_A(r) Tr\Big\{O^\dagger{\mathbf r}\cdot {\mathbf E}(R) S + S^\dagger{\mathbf r}\cdot {\mathbf E}(R) O \Big\} +g\frac{V_B(r)}{2} Tr\Big\{O^\dagger{\mathbf r}\cdot {\mathbf E}(R) O + O^\dagger{\mathbf r}\cdot {\mathbf E}(R) O \Big\} + \sum_{l=1}^{N_f}\bar{q}^l (\slashed{D}-m_q) q^l -\frac{1}{4} \tensor[]{F}{_{\mu\nu}^a} \tensor[]{F}{^{\mu\nu}_a}
\end{align}

The Wilson coefficients in this case are the potentials terms, which need to be determined by a comparison between NRQCD and pNRQCD correlation functions. For the simplest case of the static potential, i.e. $m_Q\to\infty$, we can start to carry out this matching procedure for the singlet fields at 0th order in the multipole expansion. Using the mapping of fields between NRQCD and pNRQCD $\xi^\dagger({\mathbf x}_1,t)U({\mathbf x}_1,{\mathbf y}_1,t)\chi({\mathbf y}_1,t) = S(R,r,t) + r{\mathbf r}\cdot {\mathbf E}^aO^a(R,r,t)+\ldots$ one compares a correlator that in pNRQCD contains both $S$ and $S^\dagger$, thus having a direct connection to the potential $V_s(r)$ in Eq.\eqref{pNRQCDLag}
\begin{align}
 \langle \xi^\dagger({\mathbf x}_2,t)U({\mathbf x}_2,{\mathbf y}_2,t)\chi({\mathbf y}_2,t) \chi^\dagger({\mathbf y}_1,0)U^\dagger({\mathbf x}_1,{\mathbf y}_1,0)\xi({\mathbf x}_1,0)\rangle_{\rm NRQCD} \equiv \langle S(R,r,t)S^\dagger(R,r,0)\rangle_{\rm pNRQCD}
\end{align}
The use of the straight Wilson line $U({\mathbf x}_1,{\mathbf y}_1,t)={\rm exp}[i\int_{\mathbf x_1}^{\mathbf {\mathbf y}_1} d{\mathbf z}\; {\mathbf A}({\mathbf z},t)]$ makes the above expression gauge invariant. Integrating the quadratic heavy quark part of Eq.\eqref{NRQCDLag} on the left and performing a standard Gaussian integration with the pNRQCD Lagrangian on the right, one finds that\vspace{-0.2cm}
\begin{align}
 \delta^{(3)}({\mathbf x}_1-{\mathbf x}_2)\delta^{(3)}({\mathbf y}_1-{\mathbf y}_2)\Big\langle {\rm exp}\Big[-ig\int_\square dx^\mu A_\mu\Big] \Big\rangle_{A,q,\bar{q};E<m_Q} \equiv \delta^{(3)}({\mathbf x}_1-{\mathbf x}_2)\delta^{(3)}({\mathbf y}_2-{\mathbf y}_2) {\rm exp}\Big[-itV_s(r)\Big]
\end{align}
Hence the rectangular real-time Wilson loop $W_\square(r,t)=\Big\langle {\rm exp}\Big[-ig\int_\square dx^\mu A_\mu\Big] \Big\rangle$ is intimately related to the static potential of pNRQCD. Note that the equivalence of the right- and left-hand side holds only at late times, as one replaces a field theoretical interaction with an instantaneous potential. Before elaborating on details of the computation of the values of $V_s(r)$ let us briefly look at an alternative non-relativistic description of the  of the $Q\bar{Q}$ system.

\subsection{Quantum Mechanical Path Integral: Heavy Quarks as QM Propagators}
\label{QMPathInt}

In this approach \cite{Barchielli:1986zs,Barchielli:1988zp} one does not construct a fully fledged field theory but instead embraces a non-relativistic language in terms of point particles. In particular we rely on the fact, known from the works of Feynman \cite{Feynman:1948ur}, that the propagation amplitude for a non-relativistic system can be rewritten in terms of a path integral over fluctuating trajectories $\mathbf z$ and their conjugate momenta $\mathbf p$.

Our aim is to express the propagation of heavy quarkonia in terms of such a path integral as it will allow us to read off the Hamiltonian and hence the interaction potential from within the exponential weight ${\rm exp}[i\int dt( {\bf p}\dot{{\bf z}}-H)]$, using the transfer matrix prescription. The forward $Q\bar{Q}$ correlator is chosen as a suitable quantity to act as heavy quarkonium propagator as it describes the probability to find a bare $Q\bar{Q}$ state at time $t$ when having started with a $Q\bar{Q}$ at $t=0$\vspace{-0.3cm}
\begin{align}
 D^>({\mathbf x}_1,{\mathbf y}_1,{\mathbf x}_2,{\mathbf y}_2,t) = \langle \bar{Q}({\mathbf x}_2,t)U({\mathbf x}_2,{\mathbf y}_2,t)Q({\mathbf y}_2,t)  \bar{Q}({\mathbf y}_1,0)U^\dagger({\mathbf x}_1,{\mathbf y}_1,0)Q({\mathbf x}_1,0)\rangle_{\rm QCD}\label{Eq:ForwCorr}
\end{align}
Here $U({\mathbf x},{\mathbf y},t)$ again denotes a straight spatial Wilson line, we use to make the above quantity gauge independent\footnote{One might argue that by inserting a Wilson line the question of gauge dependence has been traded for the question of path dependence. In accord with the reasoning in \cite{Aoki:2009ji} we do not expect there to be a unique potential, nevertheless all observables calculated from the different choices are expected to agree.}. As first step towards expressing Eq.\eqref{Eq:ForwCorr} quantum mechanically, we determine its form in NRQCD
\begin{align}
 D^>_{\rm NRQCD}({\mathbf x}_1,{\mathbf y}_1,{\mathbf x}_2,{\mathbf y}_2,t)=\int D[\xi,\bar{\xi}] D[\chi,\bar{\chi}] D[A,q,\bar{q}] \xi^\dagger({\mathbf x}_2,t)U\chi({\mathbf y}_2,t)  \chi^\dagger({\mathbf y}_1,0)U^\dagger\xi({\mathbf x}_1,0) e^{iS_{\rm NRQCD}}             
\end{align}
In contrast to pNRQCD we choose in the following to integrate out explicitly the quadratic heavy quark part of the NRQCD action. This Grassmann integral can be carried out analytically and amounts to replacing pairs of $\xi\xi^\dagger$ and $\chi\chi^\dagger$ with their propagators $K$ and $K^\dagger$. Since the quark fields are assumed to be nearly static, no fermionic determinant remains. Note that the difference to pNRQCD here is the focus on the propagation of each individual quark under a background field $A$ instead of a well defined singlet and octet configuration. (For a study on the propagation of a single heavy quark in the QGP see e.g. \cite{Beraudo:2010tw})

The quantities $K$ and $K^\dagger$ are the Green's functions, i.e. the functional inverses, of the NRQCD action, obeying
\begin{align}
 \Big[i\partial_t - gA_0 - m_Q +\frac{1}{2m}\Big(\partial_j+igA_j\Big)^2\Big]K({\mathbf x}_1,{\mathbf x}_2)=0, \quad \lim_{{\mathbf x}_1\to {\mathbf x}_2} K({\mathbf x}_1,{\mathbf x}_2)=\delta^{(3)}({\mathbf x}_1-{\mathbf x}_2)
\end{align} 
As such they can be reexpressed as a quantum mechanical path integral over particle paths ${\mathbf z}$ and momentum ${\mathbf p}$.\vspace{-0.2cm}
\begin{align}
 K({\mathbf x}_1,{\mathbf x}_2)=\int_{{\mathbf x}_1}^{{\mathbf x}_2} {\cal D}{\bf z}\int {\cal D}{\bf p}  \; {\cal T} \; {\rm exp}\Bigg[i\int_{0}^{t} ds \Big( {\bf p}\dot{{\bf z}} - \frac{{\bf p}^2}{2m_Q}-m_Q\Big)\Bigg] {\rm exp}\Bigg[-ig\int_{\mathbf z} dy^\mu A_\mu(y) \Bigg]\label{Eq:PathIntK}
\end{align}

Using Eq.\eqref{Eq:PathIntK} we rewrite the forward correlator in path integral language, consisting of a field independent part, containing the kinetic term and a field-dependent part, over which one still has to take the NRQCD medium average
\begin{align}
D^>_{\rm QM}=e^{-2im_Qt} &\int_{{\mathbf x}_1}^{{\mathbf x}_2} {\cal D}[{\bf z}_1,{\bf p}_1]\int_{{\mathbf y}_1}^{{\mathbf y}_2} {\cal D}[{\bf z}_2,{\bf p}_2]\;  {\rm exp}\Big[ i\sum_{l=1}^2\int_0^{t} ds \Big(  {\bf p}_l(s) \dot{{\bf z}}_l(s) - \frac{{\bf p}_l^2(s)}{2m}\Big)\Big] \left\langle {\cal T} {\rm exp} \Big[  -ig \oint dx^\mu A_\mu(x)\Big] \right\rangle
\end{align}
If the above expression is to yield a potential for the time evolution of the two-body system, we need to reexpress the last term, i.e. the real-time Wilson loop as an exponential containing the potential term of the corresponding two-body Hamiltonian. In the static limit, we assume that we can make the following identification, i.e. neglecting terms connecting different times along the path of the quark and anti-quark:  
\begin{align}
  W_\square(r,t)=\left\langle {\cal T} {\rm exp} \Big[  -ig \int_\square dx^\mu A_\mu(x)\Big] \right\rangle = {\rm exp}\Big[-i\int_0^t \; ds\; V(r=|{\bf z}_2-{\bf z}_1|,s)\Big]
\end{align}
This leads us to the defining equation for the static potential
\begin{align}
 V(r)=\lim_{t\to\infty}\frac{i\partial_t W_\square(r,t)}{W_\square(r,t)}\label{Eq:DefPot}
\end{align}

\vspace{-0.4cm}
\section{Evaluating the Potential in QCD}

Now that we have seen how a static potential can be consistently defined, starting from QCD, either along a strict effective field theory prescription (pNRQCD) or using quantum mechanical path integrals, we need to evaluate the expression Eq.\eqref{Eq:DefPot}. Perturbation theory e.g. is suited to this task as it is possible to directly calculate the forward propagator at finite temperature in real-time. Using the framework of hard-thermal loop (HTL) resummed perturbation theory, which is a gauge invariant prescription how to sum an infinite number of Feynman diagrams of the QCD medium, Laine et.al. \cite {Laine:2006ns} were the first to calculate the Wilson loop to first non-trivial order.

Their result shows that in this approximation, the static potential at temperature $T$ \vspace{-0.2cm}
\begin{align}
 V_{\rm HTL}(r)=-\frac{g^2}{3\pi}\Big[m_D+\frac{e^{-m_D r}}{r}\Big] - \frac{ig^2T}{3\pi} \phi(m_Dr),\quad \phi(x)=2\int_0^\infty dz \frac{z}{(z^2+1)^2}\Big[1-\frac{sin[zx]}{zx}\Big]\label{Eq:LainePot}
\end{align}
exhibits a Debye screened real part with Debye mass $m_D^2=g^2T^2\Big(\frac{N_c}{3}+\frac{N_f}{6}\Big)$ and an imaginary part, which can be attributed to the phenomenon of Landau damping, arising from collisions of the the light d.o.f. of the medium with the heavy $Q\bar{Q}$. 

${\rm Im}[V](r)$ at large $m_Dr\gg1$ saturates to a constant value, which can be interpreted as the energy loss of a single quark moving in a thermal bath. At small $m_Dr<1$ on the other hand the quark and anti-quark are coupled via the real part of the potential and the function $\phi(m_Dr)$ is seen to rise quadratically. The first corrections to this static potential have been calculated in \cite{Brambilla:2008cx} by going to first order in the multipole expansion in pNRQCD. The dominant contribution to the imaginary part in this scenario is identified to arise from the breakup of the singlet to an octet configuration under the influence of the color electric field.

Even though the HTL formalism is capable of reproducing thermodynamic properties of the QCD medium in good agreement with lattice QCD once temperatures are well above the critical temperature (see e.g. \cite{Burnier:2009bk}), the question remains how the medium influences the $Q\bar{Q}$ bound state at temperatures around $T_C$. Another issue is that perturbation theory is incapable of reproducing the linearly rising confining potential of the hadronic phase, hence determining whether and how this hallmark of confinement physics changes as temperatures move into the QGP regime is still an open question.

\begin{figure}[t]
\centering
 \includegraphics[scale=0.6]{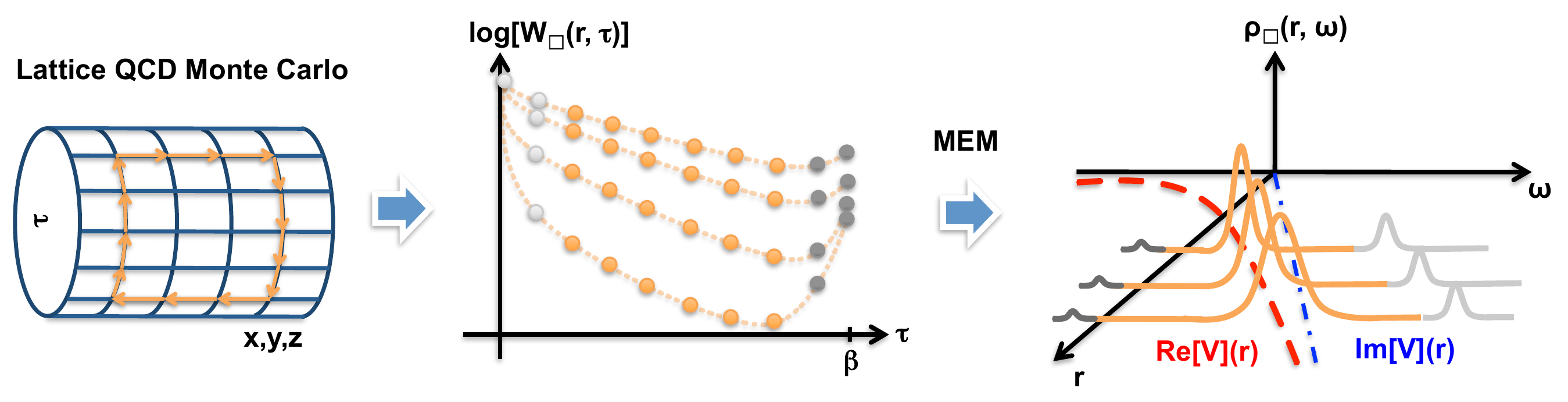}
 \caption{Schematic view of extracting the Heavy $Q\bar{Q}$ potential non-perturbatively. Starting from lattice QCD measurements of the Wilson loop, one extracts the spectra using the maximum entropy method and obtains the real and imaginary part of $V(r)$ from an inspection of the lowest lying positive frequency peak. Note that the values at $\tau=\beta$ are not related to the potential from this point of view.\vspace{-0.2cm} }
\end{figure}

Motivated by the above results, we wish to find a way how to evaluate Eq.\eqref{Eq:DefPot} non-perturbatively. One possible approach that comes to mind is lattice regularized QCD, which is amenable to Monte Carlo simulations at zero and finite temperature. A major drawback of this formalism is however that calculations are only possible in Euclidean time and it has been shown in \cite{Laine:2006ns,Beraudo:2007ky} that the corresponding Wilson loop $W_\square(r,\tau=it)$ is not directly related to the complex potential $V(r)$. We are thus facing the problem of how to analytically continue the Euclidean Wilson loop back to real-times.

We proceed by expressing the real-time Wilson loop through its Fourier transform $\rho_\square(r,\omega)$, which was shown in \cite{Rothkopf:2009pk} to be a positive definite, i.e. spectral function. As all time dependence then resides in the integral Kernel it is possible to perform the analytic continuation by merely changing from a Fourier to a Laplace transform.\vspace{-0.3cm}
\begin{align}
 W_\square(r,t)=\int_{-\infty}^\infty d\omega e^{-i\omega t} \rho_\square(r,\omega) \quad\longleftrightarrow\quad W_\square(r,\tau)=\int_{-\infty}^\infty d\omega e^{-\omega \tau} \rho_\square(r,\omega)\label{Eq:WilsonLoop}
\end{align}
The numerical problem of inverting the above relation, i.e. to extract a continuous positive definite function $\rho_\square(r,\omega)>0$ from a discrete and noisy set of Monte Carlo Wilson loop data-points, is known to be an ill-defined problem. With the help of Bayesian inference in the form of the Maximum Entropy Method, it has nevertheless become possible to give meaning to such a task\footnote{By introducing a regulator function, which supplies additional constraints to the otherwise under-determined $\chi^2$ fitting procedure, a unique spectrum is selected from the infinite set of functions that reproduce the measured data within their errorbars. Of course only parts of this spectrum are constrained by the data-points themselves, part of it is selected from the (arbitrary) choice of regulator (prior) function. Thus by variation of the prior function, one needs to identify those parts of the spectrum that stay invariant, which in turn must be fixed by the measured data.}. Details on the underlying 
concepts 
and practical challenges as well as our recent proposal for an improvement in its implementation can be found in \cite{Jarrell1996133,Asakawa:2000tr} and \cite{RothkopfMEM2011} respectively.

Once the spectrum is known, the potential can be readily extracted after inserting Eq.\eqref{Eq:WilsonLoop} into Eq.\eqref{Eq:DefPot} as shown in \cite{Rothkopf:2011db}:\vspace{-0.35cm}
\begin{align}
 V(r)=\lim_{t\to\infty}\frac{ \int_{-\infty}^\infty d\omega\; \omega\; e^{-i\omega t} \;\rho_\square(r,\omega)}{\int_{-\infty}^\infty d\omega \; e^{-i\omega t} \; \rho_\square(r,\omega)}\label{Eq:DefSpecPot}
\end{align}
The late time limit tells us that only the lowest lying peak structures\footnote{At this point a note is in order. The Euclidean Wilson loop is not a periodic quantity in the compact time direction, as the heavy quarks, which it describes, are not thermalized. Nevertheless the periodic structure of the medium gluons above $T_c$ leads to an upward trend in the Euclidean time data as $\tau\to\beta$. In the spectral picture the value of $W_\square(r,\tau=\beta)$ is thus determined by an exponentially suppressed spectral structure at negative frequencies and not by the peak responsible for the late time limit in Eq.\eqref{Eq:DefSpecPot}. (See also Fig.1)
} will play a role in the determination of $V(r)$. If e.g. there exists a single delta peak in the spectrum, the potential will be completely real and its value is given by the peak position. Similar reasoning holds for more complex peak shapes, such as a Breit-Wigner $\rho_{BW}(r,\omega)=\frac{\gamma(r)^2}{(\omega-\omega_0(r))^2+\gamma(r)^2}$, from which a complex potential ensues. Its real part being given by the peak position, its imaginary part by the peak width $V_{BW}(r)=\omega_0(r)+i\gamma(r)$. If more intricate structures arise, one needs to find improved fitting functions or in the worst case perform the full Fourier transform of Eq.\eqref{Eq:DefSpecPot}.

Our first numerical investigation of the heavy quark potential based on MEM extracted spectra of the Wilson loop in a quenched, i.e. purely gluonic medium, was reported in \cite{Rothkopf:2011db}. Using anisotropic lattices of size $20^3\times12,24,36$ at $\beta=6.1$ ($a_x=0.097fm$) and bare anisotropy of $\xi_b=3.2108$, we extracted values for both real and imaginary part at the corresponding temperatures of $T=2.33T_C,1.17T_C,0.78T_C$. In the inspection of the lowest lying positive frequency peak we fit the tip of the structure with a Breit-Wigner and read off position and width. 

The findings of our study were that below the deconfinement transition the potential is a real function and coincides with the color singlet free energies within its errorbars. As temperatures go towards and above the phase transition, the real part of the potential appears to freeze, different from the color singlet free energies, which already show screening behavior at this point. Instead we find that now a finite imaginary part accompanies the potential. 

At the highest measured temperature our results are puzzling. Both the real part and the imaginary part of the potential show a strong rise, much steeper than the T=0 potential. A priori there is no problem in the presence of such large values since the effects of real- and imaginary part might both compensate each others in the resulting time evolution, i.e. a $Q\bar{Q}$ bound by a strong real part will still melt if an equally strong imaginary part is present. Nevertheless from the point of view of asymptotic freedom such a rise is counterintuitive and we are investigating the possibility that improved fitting functions, such as a skewed Breit-Wigner, applied to the spectrum will lead to a reduction of the steep rise, bringing the real part at least down to the values found at $T=1.17T_C$. This expectation is nurtured by a comparison with the potential from Wilson line correlators in the Coulomb gauge, which show a frozen real part up to $T=2.33T_C$ while it is only the slope of the imaginary part that 
grows with temperature.

As the reliability of the MEM reconstruction of the spectra crucially depends on the quality of the data, i.e. the signal to noise ratio, it is imperative to support this first study by using larger lattices. On the one hand it has to be understood, whether the linear rise of ${\rm Re}[V](r)$ above the phase transition is really physics or related to either shortcomings of the MEM, the fitting of the spectral peaks or the quenched approximation. The same is true for the imaginary part. The MEM is known to have difficulties in reconstructing the width of peaks if the quality of data is not extremely high. Therefore improving the measurements at least above $T_C$ should be a priority. In the meantime we are performing benchmark calculations using the HTL calculated Wilson loop in real and imaginary time to compare the extracted spectra from a directly calculated Fourier transform on $W^{HTL}_\square(r,t)$ and the corresponding MEM reconstructions based on a discretized $W^{HTL}_\square(r,\tau_i)$ \cite{
BurnierRothkopf}.

\section{Heavy Quarkonium as Open Quantum System}

The finding of an imaginary part in the static $Q\bar{Q}$ potential has reignited interest in understanding the dynamics of Heavy Quarkonium melting from the viewpoint of a potential picture. In particular it is a reminder that the dynamics of heavy quarkonium need a fully dynamical treatment, as such non-hermiticities are related to the phenomenon of spatial decoherence in open quantum systems \cite{Young:2010jq,Borghini:2011yq,Akamatsu:2011se}, investigated thoroughly in condensed matter theory \cite{Breuer:2002pc}. Let us briefly sketch how the imaginary part of the static $Q\bar{Q}$ potential arises from the thermal fluctuations of the surrounding heat bath \cite{Akamatsu:2011se}.

If we assume that the full system comprised of the medium and the heavy quark system can be described in a non-relativistic language we can write down a fully hermitean overall Hamiltonian $H$ and the corresponding time evolution for the density matrix of states $\sigma(t)$ \vspace{-0.3cm}
\begin{align}
 H=H_{\rm med}\otimes I_{\rm sys} +I_{\rm med}\otimes H_{\rm sys} + H_{\rm int},\quad H^\dagger = H, \quad \frac{d}{dt}\sigma(t)=-i[H,\sigma(t)]
\end{align}
Here $H_{\rm med}$ describes the medium, $H_{\rm sys}$ the quark anti-quark pair and $H_{\rm int}$ denotes the interaction between the two. Instead of calculating in the full system, we wish to describe the physics solely in terms of the heavy $Q$ and $\bar{Q}$. I.e. we attempt to write down a Schroedinger equation with an instantaneous potential, which is supposed to encode all the interactions of the quark and anti-quark with the surrounding medium. Indeed in the derivation of the potential in section \ref{QMPathInt} we integrated out all degrees of freedom of the medium. Note that a Schroedinger equation a priori does not posses any concept of thermal fluctuations and essentially represents a vacuum equation of motion for a non-relativistic particle.

Hence in order to treat the thermal fluctuations, we consider an ensemble of wavefunctions $\Psi_{Q\bar{Q}}$ for the $Q\bar{Q}$ system and the corresponding density matrix of states\vspace{-0.2cm}
\begin{align}
 \sigma_{Q\bar{Q}}(t,\mathbf{r},\mathbf{r}')&={\rm Tr}_{\rm med}\Big[ \sigma(t,\mathbf{r},\mathbf{r}')\Big] =\langle \Psi_{Q\bar{Q}}(\mathbf{r},t) \Psi^*_{Q\bar{Q}}(\mathbf{r}',t)\rangle
\end{align}
Viewed from within the $Q\bar{Q}$ subsystem, the time evolution of the wavefunction becomes stochastic, the fluctuations encoding the interaction with the integrated out heat bath (a related approach focussing on the equation of motion for the density matrix can be found in \cite{Borghini:2011ms}). In particular it is known that these interactions with the surroundings select a certain basis of states in which the reduced density matrix $\sigma_{Q\bar{Q}}$ becomes diagonal after a time $t_{dc}$. This phenomenon is called decoherence and $t_{dc}$ is the decoherence time.

If we look back at the determination of the potential from the spectrum of the Wilson loop, a connection to such a stochastic description can be made. Instead of interpreting the width $\gamma$ of the lowest lying positive spectral peak as being directly related to an imaginary part, we can use it as an uncertainty in the value of a purely real potential. In essence this amounts to the medium gluons perturbing the color interaction between the quark and anti-quark encoded in $V(r)$ stochastically at each instant of time.

As a first attempt, we construct a model stochastic time evolution operator \vspace{-0.2cm}
\begin{align}
\Psi_{Q\bar{Q}}({\mathbf r},t)={\cal T} {\rm exp}\Bigg[ -i\int\;dt\;\Big\{ -\frac{\nabla^2}{m_Q}+2m_Q+ V({\mathbf r})+\Theta({\mathbf r},t)\Big\}\Bigg]\Psi_{Q\bar{Q}}({\mathbf r},0)\label{Eq:StochTimeEvol}
\end{align}
including a purely real potential $V({\mathbf r})$ and a real and Markovian noise term $\langle\Theta({\mathbf r},t)\rangle=0$ with non-trivial spatial correlations $\langle \Theta({\mathbf r},t)\Theta({\mathbf r',t'})\rangle=\frac{1}{\Delta t}\delta_{t,t'} \Gamma({\mathbf r},{\mathbf r}')$.
When expanding this operator according to the rules of stochastic differential calculus, we end up with the following equation of motion 
\begin{align}
 i\frac{d}{dt}\Psi_{Q\bar{Q}}({\mathbf r},t)=\Big(-\frac{\nabla^2}{m_Q}+2m_Q+V({\mathbf r})+\Theta({\mathbf r},t)-i\frac{\Delta t}{2}\Theta^2({\mathbf r},t)\Big)\Psi_{Q\bar{Q}}({\mathbf r},t)\label{Eq:StochSchroed}
\end{align}
We find that even though the time evolution in Eq.\eqref{Eq:StochTimeEvol} is fully unitary, a complex stochastic term emerges in Eq.\eqref{Eq:StochSchroed}. After the thermal average\vspace{-0.3cm}
\begin{align}
 i\frac{d}{dt}\langle\Psi_{Q\bar{Q}}({\mathbf r},t)\rangle=\Big(-\frac{\nabla^2}{m_Q}+2m_Q+V({\mathbf r})-\frac{i}{2} \Gamma({\mathbf r},{\mathbf r})\Big)\langle\Psi_{Q\bar{Q}}({\mathbf r},t)\rangle\label{Eq:StochAvgSchroedinger}
\end{align}
it leads to an imaginary part for the thermal $Q\bar{Q}$ wavefunction $\langle\Psi_{Q\bar{Q}}\rangle$. It is this averaged $\langle\Psi_{Q\bar{Q}}\rangle$, which we identify with the forward correlator calculated in the previous section. Thus the presence of the non-hermiticity of the $Q\bar{Q}$ Hamiltonian in the potential picture is directly related to the diagonal entries of the noise correlations $\Gamma({\mathbf r},{\mathbf r})$ in our model. When viewed from the level of the averaged wave function ${\rm Im}[V](r)$ appears to refer solely to a dampening phenomenon. If on the other hand we adopt the open quantum system approach, the dampening of the averaged wavefunction is only a symptom of the thermal fluctuations that reshuffle the occupancies of $Q\bar{Q}$ states during the time evolution.

Of course the above model is only a first step towards a fully open quantum systems description of heavy quarkonium in medium. Questions remain e.g. on the field theoretical derivation and correct implementation of the quantum counterpart to the classical drag force and its effects on Eq.\eqref{Eq:StochSchroed}. This part of the puzzle is crucial for the system to be able to thermalize in the late time limit. A more detailed discussion of the wavefunction dynamics and further technical aspects of the stochastic potential picture go beyond the scope of this proceeding. The interested reader is instead referred to our recent publication \cite{Akamatsu:2011se}, which also includes one-dimensional numerical simulations of the stochastic dynamics.

The author would like to thank Y. Burnier for stimulating discussions and acknowledges partial support by the BMBF under project {\em Heavy Quarks as a Bridge between Heavy Ion Collisions and QCD}.






\begin{multicols}{2}

\bibliographystyle{elsarticle-num}
\bibliography{HP2012_Rothkopf.bib}

\end{multicols}







\end{document}